\documentstyle[aps,preprint]{revtex}

\begin{document}


\vspace{2cm}
\title{Universalization as a physical guiding principle}
\author{Naresh Dadhich \footnote{email: nkd@iucaa.ernet.in}} 
\address {Inter-University Centre for Astronomy \& Astrophysics,
Post Bag 4, Pune~411~007, India}

\date{\today}

\maketitle

\begin{abstract}
In this essay, I wish to share a novel perspective based on the principle of 
universalization in arriving at the relativistic and quantum world from the 
classical world. I also delve on some insightful discussion on going 
``beyond''.

\end{abstract}
\vskip2pc 


\newpage

\section{The Newtonian World}
 
 In the classical Newtonian World, we have the absolute $3$-space 
and the absolute $1$-time with space having the Euclidean flat metric. Then 
there are the three laws of motion governing motion of particles. The laws 
however apply 
only to massive particles. That is massless particles cannot be accommodated 
in the Newtonian mechanics. If such particles exist in nature, they would 
ask for new mechanics. The characteristic of such a particle is that it 
can't exist at rest in any frame because its mass is zero. It should be 
moving relative to all observers and its speed must be the limiting speed. 
No observer can attain this speed else it could have it at rest. That is, its 
speed must be the same constant for all observers and hence it is a universal 
constant. \\

On the other hand we also know in the Newtonian framework that light always 
moves in a straight line in 
all directions. The straight line motion is indicative of constant speed. Not 
quite, there could occur accelerated motion in straight line for force acting 
along the line of motion. Since light moves straight in all directions, this 
possibility is also ruled out because there cannot exist a force acting in all directions. We are thus forced to conclude that either 
Newton's laws of motion do not apply to light or its speed must be constant. 
This constant should be the same for all observers because light moves in a 
straight line for all observers. That is, light's speed must also be a 
universal constant. \\

It is Maxwell's electrodynamics which establishes that massless particles do 
in fact exist in nature. Photons, particles of light, are indeed massless. We 
should thus have a new mechanics which incorporates massless 
particles as well in its fold. That is to incorporate a universal constant 
speed. It is the universality of laws of motion which asks for a new 
mechanics respecting universal character of the speed of light. \\

Under Newton's law of gravitation, all massive particles attract each other 
by the inverse square law and again there is no prescription for zero mass 
particles. That is, it does not incorporate massless particles in the 
gravitational interaction. Gravity is on the other hand a universal force as 
it links to all particles and ``pulls'' them all with the same acceleration  
irrespective of their mass and constitution. Linking of gravity is universal 
and hence it must also link to massless particles which however could not be 
accomplished in the Newtonian theory. Thus universality of the gravitational 
force asks for a new theory of gravitation. \\

The most fundamental universal entities are space and time and hence any 
universal property must always be expressible as a property of space and time.
 Even in the Newtonian mechanics, free particles move in straight lines which 
are geodesics (shortest distance or autoparallel curves) of the $3$-Euclidean 
space. The geodetic motion is a universal property of free particles, their 
motion does not depend upon the specific particle parameters. Also the motion 
under gravity is independent of mass of the particle. It indicates that a 
full realization of the universal character of gravitation would ask for its 
becoming a property of space and time. That is in a new theory, gravitation 
must cease to be an external field but instead be completely synthesized in 
the space-time geometry. \\  

The guiding principle that thus emerges is that a universal property must  
always be expressible as a property of spacetime. \\

\section{The Relativistic World}  

The universalization of the Newtonian mechanics which includes massless 
particles (which is in turn equivalent to incorporation of a limiting 
universal invariant speed) naturally leads to the new relativistic mechanics, 
special relativity (SR). It synthesizes space and time into a $4$-dimensional 
spacetime manifold having the flat Minkowski geometry. The universality of 
invariant light's speed gets synthesized in the definition of spacetime 
geometry specified by the Minkowski metric. \\

The next question is of universality of gravitational field. Like the laws 
of mechanics are universal (and thus must apply to all particles), so must 
be the case for gravity. It is universal and hence must act on all particles 
including particles of zero mass. The massless particles always have fixed 
speed relative to all observers. Within the Newtonian framework, action of 
force should in general change particle's speed but that can't happen for 
massless photons. We are thus faced with the real contradiction which can 
only be resolved by doing something very drastic and unusual. What is really 
required is that photon path like other particles path should instead of 
going straight bend toward the gravitating body as it grazes past it. This 
could happen without changing photon's speed only if gravitational field 
due to the body must bend space around it. That is gravity bends/curves the 
space around a gravitating body. Since space and time have 
already been synthesized, gravitational filed must hence curve spacetime. We 
thus come to a 
very revolutionary conclusion 
that universality of gravitation demands that it could honestly be described 
by nothing else but the curvature of spacetime. \\ 

 Thus universality of gravitation does not let spacetime remain inert 
background but instead impregnates it with dynamics, a unique and 
distinguishing property. It then ceases to be an external field and 
is fully integrated into the geometry of spacetime and motion under gravity 
would now be given by the free geodetic motion relative to the curved 
geometry of  
spacetime. The equation of motion for gravitational field does not have to be 
specified from outside but should by itself follow from the curvature of 
spacetime. The curvature of spacetime is given by the Riemann curvature 
tensor which satisfies the Bianchi differential identity. The contraction of 
which yields the second rank symmetric tensor, constructed from the Ricci 
tensor, having vanishing divergence. That is, 
\begin{equation} \label{div}
\nabla_{b}G^{ab} = 0 
\end{equation}
where 
\begin{equation}
G_{ab} = R_{ab} - \frac{1}{2}Rg_{ab}\,.
\end{equation}
The above equation implies 
\begin{equation} \label{gr}
G_{ab} = -\kappa T_{ab} - \Lambda g_{ab}
\end{equation}
with 
\begin{equation}
\nabla_{b}T^{ab} = 0
\end{equation}
where $T_{ab}$ is a symmetric tensor, and $\kappa$ and $\Lambda$ are 
constants. The demand that it should in the first approximation reduce to the 
Newtonian equation for gravitation will require $T_{ab}$ to represent the 
energy momentum tensor for matter and $\kappa = 8\pi G/c^2$ with $\Lambda$ 
being negligible at the stellar scale. This is the Einsteinian equation for 
gravitational field, general relativity (GR). \cite{n1,n2} \\

 Note that the constant $\Lambda$ enters into the equation naturally. It was 
first introduced by Einstein in an ad-hoc manner to have a physically 
acceptable static model of the Universe and was subsequently withdrawn when 
Friedmann found the non-static model with acceptable physical properties. 
This kind of birth gave it the name cosmological constant studded with 
ambiguity and arbitrariness about its existence. We would however like to 
maintain 
that it appears in the equation as naturally as the stress tensor $T_{ab}$ and 
hence should be considered on the same footing. What its value is to be 
determined by observations. \\

 Let us imagine, had Einstein followed the path we have done above, he would 
have concluded that space free of all removable matter/energy ($T_{ab} = 0$) 
is indeed endowed with non-trivial dynamics given by the stress tensor, 
$\Lambda g_{ab}$. This is precisely the stress tensor of virtual particles 
produced by the quantum fluctuations of vacuum \cite{weinberg}. He would have 
then anticipated that vacuum may have non-trivial quantum properties and its 
energy momentum is given by the new constant $\Lambda$. Thus $\Lambda$ comes 
in naturally as the measure of vacuum, a new constant of the Einsteinian 
gravity, a purely of general relativistic entity. This would have been a 
new prediction. Instead of $\Lambda$ being the greatest blunder of his 
life, it should have really been a profound prediction anticipating the 
quantum aspects of vacuum and its gravitational interaction. This all happens 
simply because Einstein's equation refers to the spacetime in its entirety. 
Hence whatever happens or could happen in spacetime must be contained in the 
equation. Since spacetime is all inclusive, nothing could be shielded from it.
 It is again the reflection of the universality of gravitation and spacetime. 
\\ 

\section{Another View Of Gravity}

 Let us envision that all particles massive as well as massless share a 
universal interaction without reference to gravity. That would imply that the 
interaction has to be long 
range and it can not be shielded or removed globally. Since it is shared by 
massless particles as well, the equation of motion describing motion under it 
would have to be free of mass of the particle. That would mean that it could 
however be removed locally because all particles in the local neighborhood 
would follow the same path. Then the interaction in question could 
only be removed locally but not globally. There could thus exist only local 
inertial frames (LIFs) but no global inertial frame (GIF). \\

 The non-existence of GIF and existence of only LIFs imply that spacetime has 
to be curved with its curvature being given by the Riemann curvature tensor. 
Once we have hit upon the Riemann curvature, we could follow the above line of 
derivation to get to the equation (\ref{gr}). This is now the equation of 
motion for the universal field. It is a second order non linear differential 
equation in the metric potentials, $g_{ab}$, and thus on the right it should 
have the source for the field. The second rank symmetric tensor $T_{ab}$ 
should then describe the source of the field. Since the 
interaction is universal, its source should be the property which is shared 
by all particles. That could only be energy of particles, and hence $T_{ab}$ 
should represent energy momentum tensor of matter. Then the equation becomes 
the Einstein equation for gravitational field which we have derived simply by 
using the universal character of the interaction. The envisioned universal 
interaction could therefore be nothing other than gravitation. This is quite 
remarkable that the universality property of interaction uniquely singles out 
gravity. That means gravitation is the unique universal field in the classical
 framework. \\ 

 Another example of a dynamical property determining dynamics of force is 
Bertrand's theorem for central forces in classical mechanics \cite{gold}. 
The demand of existence of closed orbits picks out two force laws, Hooke's law
 giving simple harmonic motion and the inverse square law. Further the demand 
of long range would then single out the latter. Here we refer to the most  
general property of universality of force which not only determines the 
dynamics of the force completely and uniquely but it does so in a very 
extraordinary manner by making the spacetime itself dynamic \cite{n1}. This 
happens 
because a universal feature could only be answered through a universal entity,
 spacetime. Since universality refers to a force, hence the answering 
universal agency must endow the dynamics of the force. \\   

\section{Yet Another View}

 Gravity is a classical field and hence it is a valid question to ask for 
its charge/source and the answer is matter/energy in any form. Non 
gravitational matter/energy is always positive which means it has only one 
kind of polarity and hence it can never vanish. That is a body can never be 
gravitationally neutral. From a classical standpoint, a body which is not 
charge neutral cannot be stable for it must keep on accreting until it 
neutralizes. For gravity, this can't happen for any further accretion leads 
to increasing rather than neutralizing charge. The only way this situation 
could be made sensible is that gravitational field energy must possess charge 
and it should have opposite polarity than that of non gravitational energy. 
That is why gravitational field energy must be negative and hence the field 
attractive. The field energy is however not localizable and therefore there 
is inherent ambiguity in its definition and that is why there exist several 
expressions for it in the literature \cite{berg,by}. Each of them brings out 
some aspect or the other. The main cause of ambiguity is the nonlocal or 
quasilocal character of gravitational field energy.  \\

 If we want to include gravitational field energy as a source, we have, rather 
than the usual Laplace equation for empty space, the Poisson equation with 
the field energy density on the right hand side. Its solution would now  
violate the inverse square law which is required by the conservation of charge 
for a spherically symmetric source. We thus end up in a contradiction similar 
to light's propagation in gravitational field. That is, we shouldn't violate 
the celebrated inverse square law and yet the contribution of gravitational 
field energy should be included. We have considered the equation relative to 
a fixed flat spacetime. Could it happen that if we curve the space, its 
curvature 
could account for the field energy contribution leaving the good old 
Laplace equation undisturbed and thereby the inverse square law.\\  

 The only possible way out of the contradiction is again that gravity must 
curve space (spacetime) and it can only be described by its curvature. As 
before then the Einstein equation would follow. \\

 In a simple minded fashion, we can argue as follows. Let $M_0$ be the 
bare non gravitational mass of a particle as measured at infinity with 
vanishing gravitational field energy. At any finite radius, $R$, 
in addition it would also have gravitational field energy, and so we write
\begin{equation}
M(R) = M_0 - M^2/2R
\end{equation}
which when solved for $M(R)$ gives
\begin{equation}
M(R) = -R + \sqrt{r^2 + 2M_0R}
\end{equation}
This shows that $M(R)$ vanishes as $R\rightarrow0$ and tends to $M_0$ at 
infinity. That means as $R\rightarrow0$, $M_0$ is completely eaten up by the 
negative gravitational field energy. If the particle has electric charge, 
there will remain residual mass proportional to electric charge due to 
electric field energy. This simple argument does indeed give the right 
result as obtained by the rigorous calculations \cite{adm}. The critical point 
that emerges is the fact that when gravitational field energy is included 
with the total mass and not the bare mass participating in the interaction, 
the total mass remains always finite. Even when rigorous exact calculation is 
done, this feature remains true and hence it should lead to the same result. 

 The important point to note is that the rigorous consideration does not 
invalidate this critical point and hence the result obtained here, save for 
some matter of detail, would remain true and valid on sound physical 
consideration as well \cite{by}. This is not however the case for the escape 
velocity argument for black hole and nor for the equipartition between the 
Compton length 
and the Schwarzschild radius for the Planck length.\\ 

 All this is because of the unique and bizarre property of gravitational field.
Its charge is rather a very subtle issue. It is of two kinds; one localizable 
non gravitational energy while the other non localizable field energy. It is 
the former which is like electric charge 
we know about while the latter is completely new and unique to gravity. The 
latter could honestly be described only through the space curvature \cite{n3} 
and consequently gravitational field by spacetime curvature. Now the most 
pertinent question that arises is that non gravitational matter fields can be
by prescription kept confined to, say $3$-space/brane, could we do the same 
to gravitational field? Could gravity remain confined to $3$-space or be free 
to propagate in higher dimensions as well? This is the question that refers 
to yet unexploited aspect of gravity.

 The field equation for gravitation as derived above does not refer to any 
dimension of space and would hence be valid for all dimensions. It is a truly 
universal equation. In the equation, we can prescribe to confine the matter 
stress tensor, $T_{ab}$ to the $3$-brane but $\Lambda$ refers to vacuum which 
cannot obviously be confined to any dimension. Thus gravity can not be 
confined to any dimension instead it freely propagates in all dimensions. 
This is purely a classical argument without any reference to string 
theory which requires higher dimensions. We would like to argue that even from 
classical standpoint, gravity is indeed a higher dimensional interaction. 
  
\section{The Quantum World}

 We rode on light to come to the relativistic world from the classical world. 
Once again let us ride on light. We incorporated light in the mechanics by 
treating it as a zero mass particle. Light also propagates as a wave and its 
motion is determined by the wave vector. Anything that moves must carry 
energy and momentum which are described by the $4$-momentum vector, $P_a$. 
Thus a wave, which is fully characterized by a $4$-wave vector, $k_a$ should 
also have a $4$-momentum vector associated with it describing the energy and 
momentum it carries along. There cannot exist two independent $4$-vectors 
characterizing the motion of a wave. The only sensible thing that can then 
happen is that the two vectors be proportional, $P_a = hk_a$ which would then 
readily lead to the well-known relation, $E=h\nu$ relating wave's frequency 
with the energy it carries. This should be universally true, giving the 
quantum law with $h$ being the universal constant which could be identified 
with the Planck's constant.\\ 

 The wave character of motion implies that it can not be localized in any 
frame. Non localizability essentially means uncertainty in simultaneous 
determination of the conjugate variables, position and momentum. In other way, 
it could be viewed that the process of determining one introduces uncertainty 
in the other. This happens because the energy required in the measurement 
process 
(probe's energy) is not negligible compared to the particle's energy. This 
phenomenon of uncertainty in determining the conjugate variables 
simultaneously should therefore be universal and true for all particles. We 
are thus led to the famous uncertainty relation, 
($\delta x^i\delta p^j \geq h\delta^{ij}$), which is the key relation 
of the quantum world.\\    
 
 Interestingly, it is again the incorporation of light as a zero mass 
particle or a wave in the mechanics that leads to the basic quantum principle 
and consequently to the quantum world. The quantum principle is universal and 
hence according to our general belief and the guiding principle, it should 
be expressible as a property of spacetime. This has unfortunately not 
happened despite the quantum theory being over 100 years old. Unlike SR and 
GR, quantum theory is not complete and I would believe that the completion 
would come about only when the quantum principle is expressed as a property of 
spacetime structure. We shall pick up the thread of incompleteness in the 
next section while looking beyond. \\
   
\section{And Beyond} 

 We have so far employed the principle of universalization to chart the 
passage from Newton to Einstein. The key to this course was the existence 
of zero mass particles in the nature. Their incorporation in mechanics led to
SR and their interaction with gravity led to GR. This is how we come to the 
relativistic world from the classical world. Now comes the question of going  
``beyond''. Does there still remain some unexploited aspect of 
universalization which could be tapped to go beyond? The distinguishing 
feature of the relativistic world is its anchoring on the most fundamental 
structure - spacetime. SR binds space and time while GR impregnates it with 
dynamics of gravitational field. Gravity ceases to be a force as it is 
completely synthesized in the spacetime geometry which now becomes a physical 
entity like any other force. On this line of approach, it seems quite clear 
that the way beyond could only come about by exploiting some aspect of 
spacetime which has so far remained untapped. \\

 So far we have taken spacetime to be a 4-dimensional continuum which is 
commutative;i.e. the commutator $[x^i,x^j]=0$. The commutativity and 
dimensionality of spacetime are yet free and we should keep our mind open 
about them. \\

 Universalization always lead to enlargement of the existing framework. For 
example, inclusion of light's interaction with gravity lead to the fact that 
spacetime has to be curved - a passage from flat to curved. In our 
exploration of ``beyond'' is to identify the contradiction and then attempt 
to guess the extrapolation required to resolve the issue. \\

\subsection{Gravitational world}

 In the Einsteinian world, spacetime is a $(3+1)$-dimensional manifold and the 
presence of matter/energy curves it and impregnates it with dynamics of 
gravitational field. The equation that follows from the curvature of 
spacetime has no reference to the dimension of spacetime and hence is truly 
valid in all dimensions. Secondly on the right of this equation, there are two 
terms, one the stress tensor for non gravitational matter fields and the 
other is the constant $\Lambda$ which characterizes vacuum - empty space. The 
former could be confined by prescription to a particular dimension, say 
$3$-brane but the latter is vacuum which cannot be kept confined to any 
dimension. That is,the gravitational field is however free to propagate in  
all higher dimensions even when the matter fields are confined to a $3$-brane. 
Gravity is thus truly a higher dimensional interaction and this conclusion 
follows from purely a classical consideration without reference to string 
theory \cite{gw}.\\ 

 Let us work up yet another argument for higher dimensions. A dynamics of a 
physical field cannot be fully determined unless the proper boundary 
conditions are prescribed. The dynamics of gravitational field is described 
by the curvature of 
$3$-brane, that is the $3$-space should have the proper boundary condition. 
What could that 
be? For matter fields confined to 3-brane, $\Lambda$ vacuum in the higher 
dimensional bulk space would define the proper boundary condition. The 
curvatures of the two spacetimes would be related by the well-known 
Gauss-Codazi equation. For the 
matter confined to $3$-space ($3$-brane), bulk spacetime would be 
$5$-dimensional and the source for curvature in extra dimension is only 
$\Lambda$. The universe we live in could be thought of as hypersurface or 
domain wall in the bulk spacetime. Since the matter is assumed to be confined 
on the 3-space, gravitational field which would propagate in the bulk as well 
would have $Z_{2}$-symmetry in the extra dimension. \\
 
 The picture that emerges is similar to the braneworld gravity \cite{add,rs}. 
The Gauss - Codazi equation connects the curvature tensors of bulk and brane 
spacetimes while the Israel junction conditions \cite{is} connect the 
brane extrinsic curvature with the brane tension and stress tensor. From this 
results an 
effective modified Einstein equation on the brane \cite{sms} while the bulk 
satisfies the $\Lambda$-vacuum equation (we would not here write the 
equation and the boundary conditions etc.). The equation on the brane is not 
closed 
because it has both local and non-local parts. The former arise from the 
brane stress tensor and its ``square'' while the latter from the projection
 of the bulk Weyl curvature onto the brane, which is trace-free and hence is 
called dark radiation. Further requiring that gravity though free to move 
in extra dimension should however remain localized (massless graviton to have 
ground state on the brane) to the brane where its 
source sits leads to the bulk $\Lambda < 0$, anti de Sitter spacetime 
\cite{rs,gt} (it is a different matter that it is also motivated by the 
AdS-CFT correspondence). The Newtonian potential gets modified and it has an 
additional 
$1/r^3$ term. This is what filters down as the high energy correction to the 
Newtonian gravity. The question of localization of gravity has also been 
studied for the FRW brane cosmology and it requires $\Lambda$ on the brane to 
be non-negative \cite{rk,pd}. If localization of gravity is taken as the 
determining factor for the FRW cosmology, non-negative $\Lambda$ is a definite
 prediction of the braneworld gravity. \\

 We would thus like to say that braneworld picture of gravitation naturally 
arises simply from the universal property. This is quite independent and 
separate motivation. It is the universality property which is responsible 
for gravity being a higher dimensional interaction. We have also enunciated 
an interesting and novel argument for the self interaction of gravitational 
field and it being attractive simply by demanding stability of a gravitating 
system.\\ 

\subsection{Quantum gravitational world}

 In our discussion we have so far come across four physical constants, 
$c, h, G$ and $\Lambda$. All of them are different in character and signify 
different physical aspects. They are as follows: \\

$\bullet \, c$ is the most fundamental of the constants. Recently its six different 
aspects have been enunciated \cite{ellis} showing its various avatars. The 
most profound amongst them is that it represents a symmetry of spacetime. 
It is a universal spacetime property. None of the others has this 
distinction. \\

$\bullet \, h$ is also represents a universal physical principle but like $c$ it has 
not yet been synthesized into the structure of spacetime. Anything that is 
universal should be expressible as a spacetime property. This is the main 
problem of the quantum theory and we yet have no clue about the solution. \\
 
 $\bullet \, G$ is simply the coupling constant of spacetime geometry 
(gravitational field) with the non  gravitational matter fields. It is really 
the measure of the universal aspect of matter fields which links to spacetime 
geometry. It is the coupling constant for the universal field in which all 
matter fields participate. The interaction does indeed become a property 
 of spacetime and it is described by its curvature. \\

 $\bullet \, \Lambda$ is the new 
constant of the Einsteinian gravity. It is the measure of vacuum's coupling to 
spacetime geometry. Vacuum is by definition universal and hence $\Lambda$ is 
indeed a new universal constant. Unlike all others, there prevails a profound 
ambiguity about its character as a true universal constant particularly 
because of the chequered history of its introduction and subsequent uncertain 
and ambiguous existence. In our derivation of the field equation for 
gravitation, it follows
naturally and is in fact on the same footing as the stress tensor for matter 
field. It is indeed the stress tensor of vacuum. This is the new prediction of 
the Einsteinian gravity which anticipates the stress tensor of vacuum 
fluctuations.\\

 $\Lambda$ is truly a measure of what could happen to vacuum. At the micro 
level it connects to the quantum fluctuations while at the large scale it 
defines the size of the Universe - rather it provides a closure to the 
Universe. It could be viewed as defining the underlying spacetime substratum 
as the matter free limit of the universe. The remarkable fact is that it is 
non trivial and has dynamics which is entirely described by $\Lambda$.\\

 Dual role of things is quite characteristic of GR. For instance, 
gravitational potential in the Schwarzschild solution does the dual role of 
giving the Newtonian inverse square law as well as curving the space for 
photons to bend. How could $\Lambda$, which is a pure GR creature, therefore 
escape this distinguishing feature? At one end it could be a measure of 
quantum vacuum fluctuations while at the other it could source the cosmic 
acceleration. 
Could it bind the smallest scale of the Universe with the largest and obtain 
a profound synthesis? This is the most intriguing and interesting question.\\

 We therefore believe that any attempt to quantum theory of gravity must 
have to address to $\Lambda$ in a non trivial manner because it is a true new 
constant of the relativistic gravity \cite{gp}. On the other hand, 
almost all the attempts peg on the Planck length as the most fundamental 
quantity. The Planck length is a construct made out of the three constants 
and it does not follow from any theory or physical principle. Also note that  
the three constants that make the Planck length are all not on 
the same footing, for instance 
$c$ represents the Lorentz symmetry of spacetime while none other has such a 
fundamental claim. We would argue that a constant that truly follows from 
a theory should always have precedence over a construct and hence has to 
be first addressed.\\

 The only physical argument being advanced for the Planck length is 
that it represents the equipartition between the Compton wave length and the 
Schwarzschild radius. All this presumes that the whole framework which defines 
these lengths remain valid at the energy which is 16 orders of magnitude 
higher than anything we have known so far. This comparison is supposed to be 
done against a fixed background, how could that remain valid at such a high 
energy - spacetime has to be curved at such high energy? This simplistic 
argument might give the right answer though based on 
wrong reasoning. We have the famous example of this kind. 
Laplace and Michell obtained the exact Schwarzschild radius by computing the 
escape velocity for a photon. We all know that it is wrong.\\ 

 Let me add one more to this kind of tentative consideration by comparing 
$\Lambda$ with the Schwarzschild radius and we obtain the mass scale, 
$c^2/{G\sqrt{\Lambda}}$ of the order of $10^{57}$ grams equivalent to 
$10^{80}$ nucleons, the total mass of the Universe. In the relativistic 
world, we have the three constants which yield to this measure of mass. This 
is a valid scale of the relativistic framework. \\

 My objection to these considerations is not only because they are tentative 
and based on invalid premise but more so because they blind one from the real 
insight. If you accept the escape velocity argument for photon, it closes the 
inquiry as there is nothing further to inquire into. While its non acceptance 
should have raised the question whose solution would have naturally ended up 
in a new theory of gravity, GR. To me the Planck length should have the 
similar kind of blinding effect and it is perhaps hiding the crucial question 
leading to a correct theory of quantum gravity.\\

 Let us again follow our natural lead of identifying the contradiction in the 
existing framework and then seeking its resolution by enlarging  the 
framework. For quantum gravity, the contradiction is, matter, which produces 
gravity, has at high energy quantum behaviour while the spacetime which 
describes gravity is inherently continuum. Thus arises the contradiction. How 
do we bypass it? \\

 Going by our earlier experience, we recall that for accommodating light's 
motion in mechanics we had to bind space and time into a synthetic whole 
spacetime, and for light to feel gravity we had to curve spacetime. Now what 
could we do to spacetime to accommodate quantum behaviour of matter which 
curves it? We had so far not altered the basic structure of space and time, 
and have always managed by adding a new property. Could we still invoke 
some new property of spacetime which lets us resolve the contradiction? What 
is required is to make spacetime curvature discrete. Could curvature be 
made quantum yet specetime retaining its continuum character? This is the 
first question. If yes, that would indeed be very good and comfortable. 
Otherwise we are left with no other option but to make spacetime itself 
quantum which would obviously be a very formidable task. The loop quantum 
gravity precisely does that \cite{lqg}. \\

{\it Frankly I am totally confused here.}

 Let us note that the equation for gravitational field has as we have 
obtained two components on the right, one stress energy tensor of matter 
and the other of vacuum. The former certainly has the quantum structure 
at high energy while this will not be the case for the latter. Secondly, the 
former could be 
confined to a given dimension, say the 3-brane, while the latter cannot be 
and hence we have to consider gravity as essentially a higher dimensional 
field. 
When we are attempting to quantize gravity, we are in totally 
new and unfamiliar territory. So far we have done quantization of fields 
and matter against a fixed spacetime background. Now we have to quantize 
spacetime against itself. If this is not crazy, it is certainly quite strange 
and bizarre. Now we have an added new feature that the higher dimensional 
vacuum would continue to have the continuum structure. Could it then happen 
that we should consider quantization of gravity (curvature of 3-brane) in the 
continuum spacetime of higher dimensional vacuum (anti/de Sitter) spacetime. 
I am not sure how to implement it but it certainly opens up a new window 
which is in line with what we have done earlier. \\

 Another way of looking at quantization is to ask for synthesis of $h$ with 
the spacetime structure.  This is because quantum 
principle is universal and hence it must be expressible like $c$ as a property
of spacetime. This will render quantum theory complete on par with the 
relativity theories, SR and GR. The main feature of the quantum framework is 
non localizability leading to non locality. Note that incorporation of 
higher dimensional feature of gravity would naturally lead, as in the 
braneworld gravity, to non local effects. Perhaps it is this non locality 
which could be exploited or synthesized with the quantization procedure. 
Could this make space non commutative which may facilitate quantum 
behaviour? Howsoever crazy may it all look yet, since we are faced with the 
most formidable and complex problem, no stone should be left unturned.  
It may therefore happen that synthesis of $h$ into spacetime also requires 
$G$ along. That is quantum principle could not be synthesized into spacetime 
alone but it requires gravity along as an inseparable partner. In other 
words, the quantum theory cannot be completed in flat spacetime, spacetime 
has necessarily to be dynamic and hence curved. This seems quite natural for 
the synthesis being sought could only occur at very high energy where 
spacetime has to be dynamic and curved. \\

 Theory of quantum gravity/spacetime is therefore required for completion of 
quantum theory as well and which would have to include gravity along. \\

 So far almost all the attempts to quantum gravity have been anchored on the 
Planck length. In the string theory, 
quantization is done in higher dimensional flat space and GR together with 
other fields appear as the low energy effective theory in $4$ dimensions. We 
would rather argue  
that higher dimensions cannot be flat, they should at least contain $\Lambda$ 
hence string theory should manipulate at least the anti-de Sitter space. The 
background dependence of the programme is being recognized as one of the 
major weaknesses of the string theory approach. On  
the other hand the loop quantum gravity programme does however deal with 
curved spacetime but restricted to $4$-dimensions \cite{lqg}. We believe that 
this programme should in some manner have to be supplemented to incorporate 
higher dimensional nature of gravity. \\

 So far the two main approaches to quantum gravity have remained quite
separate and distant from each-other. It is perhaps time that the two should  
tend to converge. They perhaps address the complimentary aspects of gravity 
in their framework and asymptotically we should be able to understand the 
relation between them.\\

 Going by our adherence to universalization and its proper expression in 
spacetime property as the abiding guiding principle, we argue that gravity is 
intrinsically higher dimensional 
field as much as it could only be described by curvature of spacetime. While 
its basic source, matter fields can remain confined to $3$-brane but the 
field can propagate off the brane. The dynamics of the field in bulk is 
solely determined by $\Lambda$ and spacetime there can continue to have the 
usual continuum character. It has to be quantum on the brane. As shown for 
the braneworld gravity, the field on the brane has non local aspect. Non 
locality is a distinguishing feature of quantum theory. Could this 
incorporation of non locality on the brane facilitate quantization of gravity? 
On the face of it, this approach would actively involve both $\Lambda$ and 
$h$ along with of course $c$ and $G$. This is precisely what we had asked 
for that both $\Lambda$ and $h$ should be involved in quantum gravity. The 
only other aspect of spacetime which remains still free is commutativity. 
We could bring in non commutativity if required. This is just a suggestion 
which is prompted by our general principle of universalization which has so 
far been reliable and insightful. \\

 In the above I have argued hopefully convincingly the following points:\\
 
(a) $\Lambda$ is not only a true new constant of the relativistic gravity but 
is the measure of the dynamics of vacuum and which cannot be confined to 
any specific dimension.\\

(b) We have also shown that gravity is the unique universal field. We have 
derived the Einstein equation simply appealing to the universal character 
of the field. \\

(c) In an innovative way we have shown why gravity is attractive and is 
inherently higher dimensional and could only be described by the 
curvature of spacetime. \\
 
(d) It is the completion of the quantum theory which naturally asks for a 
quantum theory of spacetime and hence of quantum gravity. The synthesis 
of quantum with spacetime could only occur when gravity is also included. 
That is spacetime can attain quantum structure only at high energy which 
would necessarily render spacetime dynamic. For gravity, $\Lambda$ is 
unshakable and hence quantum gravity has to in some way involve both $h$ and 
$\Lambda$. \\

(e) We make a new proposal which takes into account higher dimensional 
character of gravity and also the fact that vacuum part (bulk spacetime) could 
continue to have the continuum structure. Synthesizing this with the quantum 
character of the $3$-brane space is the challenge. This approach would 
involve both $h$ and $\Lambda$ quite naturally. This is the feature which we 
had set out for a theory of quantum gravity. \\  

 I hope that the above discussion would perhaps open up a new perspective 
and insight which may turn out to be illuminating.  

{\bf Acknowledgment:} This perspective has evolved over a period of time and 
I was set on pooling together all my thoughts in a coherent manner while 
preparing for the two lectures \cite{n1,n2} and hence the organizers of these 
two events deserve my thanks. I would also like to thank Parampreet Singh and 
Mohamad Sami for acting as a good sounding board.


\begin{references}
\bibitem{n1} N. Dadhich, Subtle is the Gravity, gr-qc/0102009.
\bibitem{n2} N. Dadhich, The Relativistic World: A Common Sense Perspective, 
physics/0203004.
\bibitem{weinberg} S. Weinberg, Rev. Mod. Phys. {\bf 61}, 1 (1989).
\bibitem{gold} H. Goldstein, {\it Classical Mechanics}, Addison-Wesley (1980). 
\bibitem{berg} G. Bergqvist, Class. Quantum Grav. {\bf 9}, 1753 (1992).
\bibitem{by} J. D. Brown, J. W. York, Phys. Rev. {\bf D47}, 1407 (1993).
\bibitem{adm} R. Arnowitt, S. Deser, C. W. Misner, in Gravitation: an 
introduction to current research, John Wiley (1962), p.227.
\bibitem{n3} N. Dadhich, On the Schwarzschild Field, gr-qc/9704068.
\bibitem{gw} M. B. Green, J. H. Schwarz, E. Witten, {\it Superstring 
Theory},  Cambridge University Press (1988); J. Polchinski, {\it String Theory}, Cambridge University Press (1998).
\bibitem{add} N. Arkani-Hamed, S. Dimopolous, G. Dvali, Phys. Lett.
{\bf B 429}, 263 (1998).
\bibitem{rs}L.~Randall and R.~Sundrum, Phys. Rev. Lett. {\bf 83}, 4690 (1999).
\bibitem{is}W. Israel, Nuov. Cim. {\bf B 44}, 1 (1966).
\bibitem{sms}T. Shiromizu, K. Maeda, M. Sasaki, Phys. Rev.  {\bf D 62}, 024012 (2000).
\bibitem{gt} J. Garriga, T. Tanaka, Phys. Rev. Lett. {\bf 84}, 2778 (2000).
\bibitem{rk} A. Karch, L. Randall, JHEP {\bf 0105}, 008 (2001).
\bibitem{pd} P. Singh, N. Dadhich, ``Localization of gravity in brane world cosmologies'', hep-th/0204190 (To appear in Mod. Phys. Lett. A);
``Localized gravity on FRW branes,'' hep-th/0208080.
\bibitem{gp} R. Gambini, J. Pullin, Class. Quant. Grav. {\bf 17}, 4515 (2000).
\bibitem{ellis} G. F. R. Ellis, J-P. Uzan, 'c' is the speed of light, 
isn't it?, gr-qc/0305099.
\bibitem{lqg} C. Rovelli, {\it Loop Quantum Gravity}, Living Reviews, {\bf 1}, \\
http://www.livingreviews.org/articles; T. Thiemann, ``Introduction to Modern 
Canonical Quantum General Relativity,'' gr-qc/0110034.


\end{references}
\end{document}